\documentstyle[12pt]{article}

\def\fun#1#2{\lower3.6pt\vbox{\baselineskip0pt\lineskip.9pt
        \ialign{$\mathsurround=0pt#1\hfill##\hfil$\crcr#2\crcr\sim\crcr}}}

\renewcommand\({\left(}
\renewcommand\){\right)}
\renewcommand\[{\left[}

\newcommand\eq[1]{Eq.~(\ref{#1})}
\newcommand\eqs[2]{Eqs.~(\ref{#1}) and (\ref{#2})}

\newcommand\pa{\partial}

\newcommand\ee{\end{equation}}
\newcommand\be{\begin{equation}}
\newcommand\eea{\end{eqnarray}}
\newcommand\bea{\begin{eqnarray}}

\newcommand\TeV{\,\mbox{TeV}}
\newcommand\GeV{\,\mbox{GeV}}

\newcommand\eV{\,\mbox{eV}}

\newcommand\mpl{M_{\rm P}}

\newcommand\lsim{\mathrel{\rlap{\lower4pt\hbox{\hskip1pt$\sim$}}
    \raise1pt\hbox{$<$}}}
\newcommand\gsim{\mathrel{\rlap{\lower4pt\hbox{\hskip1pt$\sim$}}
    \raise1pt\hbox{$>$}}}

\def\dslash{\not{\hbox{\kern-2pt $\partial$}}}
\def\Dslash{\not{\hbox{\kern-4pt $D$}}}
\def\Oslash{\not{\hbox{\kern-4pt $O$}}}
\def\Qslash{\not{\hbox{\kern-4pt $Q$}}}
\def\pslash{\not{\hbox{\kern-2.3pt $p$}}}
\def\kslash{\not{\hbox{\kern-2.3pt $k$}}}
\def\qslash{\not{\hbox{\kern-2.3pt $q$}}}

 \newtoks\slashfraction
 \slashfraction={.13}
 \def\slash#1{\setbox0\hbox{$ #1 $}
 \setbox0\hbox to \the\slashfraction\wd0{\hss \box0}/\box0 }
 
\def\ee{\end{equation}}
\def\be{\begin{equation}}

\newcommand\sub[1]{_{\rm #1}}

\begin{document}
\begin{titlepage}
\pagestyle{empty}
\begin{center}
\hfill LBNL--42536\\
\hfill LANCS--TH/9821\\ 
\hfill hep-ph/9811375 \\
\hfill November 1998
\vskip .3in
{\large \bf Quintessential difficulties}\footnote{
This work was supported in part by 
NATO grant CRG 970214 and US Department of Energy contract 
DE--AC03--76SF00098.}
\vskip .2in
\vskip .2in
Christopher Kolda$^{\dag}$ and David H. Lyth{}$^{\ddag}$ 
\vskip .2in

{}$^{\dag}${\em Theoretical Physics Group, Lawrence
 Berkeley National Laboratory, \\ University of California, 
 Berkeley, California 94720, U.S.A. }\\[2mm]

{}$^{\ddag}${\em Department of Physics, Lancaster University, 
Lancaster, LA1 4YB, U.K.}

\end{center}

\vspace{.6cm}
\begin{abstract}

An alternative to a cosmological constant is quintessence, defined as a
slowly-varying scalar field potential $V(\phi)$. If quintessence is 
observationally significant, an epoch of inflation is beginning at the
present epoch, with $\phi$ the slowly-rolling inflaton field. In 
contrast with ordinary inflation, quintessence seems to require extreme fine
tuning of the potential $V(\phi)$. The degree of fine-tuning is 
quantified in various cases.

\end{abstract}
\end{titlepage}

{\bf 1.} In the context of Einstein gravity, a cosmological constant may
be regarded as a constant contribution to the energy density of the 
Universe. One can instead consider a contribution, termed quintessence
\cite{quint1,gb,quint,carroll,bin},
 which is slowly decreasing on the Hubble timescale at the present 
epoch, and will presumably vanish in the infinite future.
The variation in one Hubble time might be negligible, in which case
quintessence is observationally the same as a cosmological constant,
or it might be significant.

There is evidence, not yet compelling, that a cosmological constant
or quintessence gives a significant contribution to the 
present energy  density,
with some leaning towards the latter
\cite{obs}. Such a contribution is of order 
$\sim (10^{-3}\eV)^4$,
the present value of 
the critical energy density
$3\mpl^2H^2$,
and for the sake of simplicity one assumes
that the total energy density has the critical value.
(As usual $\mpl\equiv(8\pi G)^{-1/2}=2.4\times 10^{18}\GeV$ is the 
reduced Planck scale, and $H$ is the Hubble parameter.)

A cosmological constant may be regarded as a nonzero value of the
effective scalar field potential $V$, at the minimum which corresponds 
to our vacuum. From a theoretical viewpoint, it is not clear how
the required value $V\sub{vac}\sim (10^{-3}\eV)^4$ would be determined.
In units
of $\mpl$ the required value is 
\be
\frac{V\sub{vac}}{\mpl^4} \sim 10^{-120} \,.
\label{120} \ee

Quintessence corresponds to $V\sub{vac}=0$, which may be easier to 
understand. At the present epoch, $V$ is slowly decreasing towards this 
value. 
Quintessence, representing a significant fraction of the present energy 
density, is generated if the present epoch represents
the beginning of an era of inflation, with some
quintessence field $\phi$ satisfying the 
slow-roll
approximation $3H\dot\phi=-V'$, and the potential obeying 
the flatness conditions
\bea
\mpl|V'/V| &\ll& \label{flat1} 1 \\
\mpl^2|V''/V| &\ll& 1 \label{flat2}  \,.
\eea
The first condition ensures that $V$ is indeed slowly varying
on the Hubble timescale, and the second condition is required for 
consistency of the slow-roll approximation. Conversely,
with both flatness conditions satisfied, slow-roll typically represents 
an attractor for a wide range of initial conditions.

The flatness requirements \eqs{flat1}{flat2}
are usually considered in the context of the era of 
inflation that is supposed to set initial conditions for the Hot Big
Bang, which we shall call ordinary inflation.
Ordinary inflation can be achieved without any significant 
fine-tuning \cite{p97toni}.
One might therefore suppose that quintessence 
can also be achieved without extreme fine-tuning, and 
some of the literature seems to support this view.

In particular, there is the proposal  \cite{quint,bin} that
\be
V = \frac{\Lambda^{4+\alpha}}{\phi^\alpha} \,,
\label{dsb}
\ee
with $\alpha>0$ usually of order 1, 
and $\Lambda$ some mass scale.
This potential satisfies the flatness conditions at $\phi\gsim
\alpha\mpl$, and it is supposed that our epoch corresponds to the
beginning of this regime, $\phi\sim \alpha\mpl$. For $\alpha
=2$, \eq{120} is satisfied for $\Lambda \sim 1\GeV$, with larger $\Lambda$ 
for larger $\alpha$. Such values can be naturally generated by strong
coupling effects and/or dynamical 
symmetry breaking, which give $\Lambda\sim e^{-8\pi^2/bg^2}\mpl$,
where $g\sim{\cal O}(1)$ and the $\beta$-function, $b$, will usually 
be roughly 1 to 10. One seems indeed to have avoided fine-tuning.

The problem, with this or any other model of quintessence, is to
prevent additional terms in the potential $V(\phi)$ which would violate the
flatness conditions. In this note, 
we consider both the non-supersymmetric and supersymmetric cases,
with emphasis on the latter.
We focus mainly on the tree-level 
contributions to $V(\phi)$, involving the mass and self-couplings of 
$\phi$, and argue that these parameters have to be extremely small compared 
with their natural values in order to satisfy the flatness conditions.
Loop corrections to $V(\phi)$ involve also the 
couplings of $\phi$ to other fields, 
as well as their masses and (at higher order) self-couplings,
but in the supersymmetric case
we shall not attempt to quantify the degree of suppression of 
those parameters that is implied by the flatness conditions.

Our work is complementary to that of Carroll \cite{carroll}.
Instead of the flatness conditions, he discussed the constraint implied 
by the observational limits on a fifth force, in a non-supersymmetric
context. The conclusion there is that one requires a
moderate suppression of certain non-renormalizable 
couplings of the quintessence field to other fields.

{\bf 2.} Barring accidental cancellations, the flatness conditions are 
certainly going to require some internal symmetry. We shall first 
consider the case that $\phi$ is the modulus 
of some complex field, that is charged under a symmetry
acting on the phase so that $\phi=0$ is the fixed point.
This eliminates the linear term in $V(\phi)$, and for simplicity we 
assume that it eliminates the cubic term as well. 
Then the potential has the form
\be
V=V_0 + \frac12 m^2\phi^2 + \lambda_4 \phi^4 +
\sum_{d=5}^\infty \lambda_d \mpl^{4-d} \phi^d + \cdots \,.
\label{tree}
\ee
The exhibited terms correspond to the tree-level potential,
consisting of the mass term, the renormalizable quartic term,
and the non-renormalizable terms with $d\geq 5$.
The 
terms represented by dots represent quantum corrections.
The latter include loop corrections, and possible 
non-perturbative effects giving terms like the one in \eq{dsb}.

As we are trying to see whether fine-tuning can be avoided, we
discount the possibility of accidental cancellations between different 
contributions to the slope of $V$.
Then the first flatness condition
$\mpl|V'/V|\ll 1$ is implied by the second flatness condition 
$\mpl^2|V''/V|\ll 1$, unless $\phi$ is far bigger than $\mpl$.
We shall soon see that the latter regime is completely unviable,
so we need consider only the second flatness condition which gives
\bea
m^2 & \ll& V\sub{vac}/\mpl^2 \,\,\approx\,\, (10^{-42}\GeV)^2 \label{mconst} \\
\lambda_d &\ll&  \frac{V\sub{vac}}{\mpl^4} \(\frac\mpl\phi \) ^{d-2}
\approx\,\, 10^{-120}\left(\frac\mpl\phi\right)^{d-2} 
\label{lamconst}\,.
\eea

In the absence of supersymmetry, the mass is unstable
against radiative corrections.
Suppose $\phi$ couples to some other field in the theory with (dimensionless)
coupling $\zeta$. Loop corrections to the $\phi$-propagator will then shift 
$m^2$ by an amount $+\zeta\mpl^2$ if the field is a boson,
and by $-\zeta^2\mpl^2$ if it is a fermion.  
Supersymmetry ensures that the bosonic and fermionic contributions 
cancel, but without supersymmetry there is no reason for a 
cancellation. In the absence of a cancellation, the 
bound \eq{mconst} 
requires $\zeta \ll V/\mpl^4\sim10^{-120}$ for the bosonic couplings,
and $\zeta^2 \ll V/\mpl^4\sim10^{-120}$ for the fermionic couplings.
This is the same amount of fine-tuning that is needed simply to impose
the observational value \eq{120}.

Without supersymmetry, the same degree of fine-tuning is required
by the constraint \eq{lamconst} on the 
non-renormalizable couplings, unless $\phi$ is well below $\mpl$.
Indeed, the expected values of the $\lambda_d$ are of order
1 since they represent quantum gravity effects at the Planck scale.
(At least this should be the case for $d$ not too large;
for extremely large $d$ one might reasonably expect \cite{km}
a behavior like $\lambda_d \sim 1/d!$).

{\bf 3.} Henceforth, we present our discussion in the context of
supersymmetry. This ensures a cancellation between the fermionic
and bosonic quantum corrections to $m^2$, and as we shall see it gives
some control over the couplings $\lambda_d$.
Supersymmetry is treated in several texts, and a
summary of the aspects relevant for inflation is given in
\cite{p97toni}. 

Taking the usual chiral formulation, supersymmetry works with
complex scalar fields that we shall denote by $\Phi_n$. 
As a function of these fields, the
tree-level potential has a well-known form, consisting of an $F$-term
plus a $D$-term. The $F$-term 
involves the superpotential $W$, which is
holomorphic in the complex fields, 
and the real K\"ahler potential $K$ which is taken to be a function 
of the fields and their complex conjugates. 
The $D$-term involves the holomorphic gauge kinetic function
$f$ and also $K$, but it 
is unlikely to be relevant for quintessence and we ignore it for the 
moment.

Because $W$ is holomorphic, its form is very strongly 
constrained by internal symmetries. As a result, one 
can write down a simple expression corresponding to a model of 
quintessence (or anything else) and forbid all additional terms.
Because of the specific form of $V$, this gives considerable control
over $m$, $\lambda$ and the non-renormalizable coefficients
$\lambda_d$, and in particular allows one to suppress the latter
far below the generic value $\lambda_d\sim 1$.
However, in contrast with the case for ordinary inflation, this 
suppression is  nowhere near enough to make quintessence viable.

The problem arises because supersymmetry must be broken if it is to 
realized at all in nature. The scale of
supersymmetry breaking $M_S$ is very large compared with $V^{1/4}
\sim 10^{-3}\eV$. Indeed, to have a viable low-energy phenomenology
one needs $M_S\gsim 1\TeV$, 
and it is usually supposed that $M_S\sim 10^{10}\GeV$.
Also, sensible models seem to require 
at least a significant fraction of $M_S$ to come from the $F$-term.
Assuming for simplicity that $M_S$ comes entirely from the $F$-term,
and involves only say $\Phi_1$, 
the potential $V(\phi)$ in the presence of supersymmetry breaking 
is of the form
\be
V(\phi) =M_S^4 \( k(\phi)  + \cdots \) + \cdots \,,
\ee
where $k$ is the 1-1 element of 
the matrix inverse of $\pa^2 K/\pa \Phi_n
\pa \Phi_m^*$. (Contributions to $V$ can also come from $W$, but in
specific models holomorphy will often forbid such terms.)
Because it is not derived from a holomorphic
function, $k$ cannot be controlled by symmetries
acting on the phases of the $\Phi_n$. It will therefore have
an expansion
\be
k = 1 + \sum_{d=2}^\infty k_d \mpl^{-d} \phi^d \,,
\ee
with $|k_d|\sim 1$.
This will gives contributions to the mass and couplings of order
\bea
m^2 &\sim& M_S^4/\mpl^2 \label{msquared} \\
\lambda_d &\sim & M_S^4/\mpl^4 \label{lamd} \,.
\eea
The mass-squared  is a factor
\be
M_S^4/V\sim (1\TeV/10^{-3}\eV)^4 \sim 10^{60} 
\label{mass}
\ee
too big, and the same is true of the couplings unless
$\phi$ is far below $\mpl$. This represents severe fine-tuning.

Of course, it is always possible that the term in $V$ proportional to
$M_S^4$ might be suppressed because $K$ and $W$ have special forms.
This occurs in the 
form of supersymmetry
termed `no-scale', where the term actually vanishes at tree level.
But no-scale supersymmetry does not seem to 
emerge from string
 theory.\footnote
{Let us mention the two popular examples. In weakly coupled heterotic 
string theory, no-scale supersymmetry corresponds to the case that the 
superpotential $W$ is independent of the bulk moduli $t_I$, {\it i.e.},
$\pa W/\pa t_I=0$. In the true
vacuum, $W$ is non-vanishing, and because of modular invariance
one is unlikely to have $\pa W/\pa t_I=0$. 
(In contrast, for ordinary inflation
a potential of the no-scale form
can be obtained \cite{p97toni}, since 
the condition $V=M_S^4$ corresponds to $W=0$ making it easy to 
achieve $\pa W/\pa t_I=0$ without violating modular invariance.)
In Horava-Witten
M-theory, no-scale supersymmetry does not seem to emerge at all.
Finally, one might mention that a recent proposal \cite{lisasundrum}
eliminates the tree-level contribution to $m^2$, but does not suppress 
non-renormalizable interactions with the visible sector, so that 
$m^2\sim M_W^4/\mpl^2$ generically, which is still much too large.}
At present, no mechanism is known that would suppress the mass
and coefficients below the level of \eqs{msquared}{lamd}.

It might at first appear that models of dynamical SUSY breaking or
models in which exact superpotentials are calculable, such
as those employed in \cite{bin}, might work as models of quintessence
since in the large field limit $W$ is calculable and its flat
directions appear to be truly flat. While this it true, such models cannot
provide quintessence in a universe that looks like ours. In our
universe, SUSY is badly broken and that breaking is (generically)
communicated to all fields in the theory. In general, only scalars
which are already protected from receiving mass contributions ({\it e.g.},
Goldstone bosons) remain massless after SUSY-breaking.

Let us comment briefly on the possibility of constructing a 
quintessence model with $\phi\ll\mpl$, which might sufficiently suppress 
the quartic and non-renormalizable terms.
In this case,
$V$ needs to be dominated by the constant term $V_0$, because
no single term of the varying part of $V$
will satisfy the flatness conditions on its own
and we are trying to avoid delicate
cancellations. 
Given the assumption that $V$ vanishes in the true vacuum
(achieved in the far future), $V_0$ will be a function of the 
other 
parameters in the potential, but the problem will be to explain its
smallness. This difficulty explains, no doubt, why the literature does 
not contain any models of quintessence with $\phi\ll\mpl$.

{\bf 5.} We now turn to models of quintessence \cite{gb}
in which $\phi$
is a pseudo-Goldstone boson. This corresponds to an 
approximate
global $U(1)$ symmetry 
$\phi\to\phi\,+\,$const, and $V(\phi)$ is flat in the limit
of exact symmetry. We focus on the usual case, that
$\phi$ corresponds to the phase of a complex field $\Phi$,
which is in the bottom of a
Mexican Hat potential
\be
V=\lambda(|\Phi|^2 - \mu^2/2 )^2 + \cdots \,.
\ee
At the bottom of the Mexican Hat we write 
$\Phi= (\mu/\sqrt{2}) \exp(i\phi/\mu)$. 
The dots represent non-renormalizable terms and quantum
corrections which may generate a potential for $\phi$.
For a model of quintessence
(or ordinary inflation) it is  convenient to set 
$\phi=0$ at a maximum of the potential, near which inflation takes place.

In the limit of exact symmetry, $V(\phi)$ is perfectly flat.
If the global $U(1)$ is explicitly broken to $Z_N$, a potential for
$\phi$ is generated of the form
\be
V(\phi) = \frac12 V_0 [\cos(N\phi/\mu)+1]
=V_0 -\frac12m^2\phi^2 +\cdots 
\ee
where $m^2=\frac12N^2 V_0/\mu^2$. 
Proposals exist \cite{schizkim} for obtaining the required value
$V_0\sim (10^{-3}\eV)^4$, but we still have to satisfy the flatness
conditions, in particular \eq{mconst}. This requires
$\mu\gg\mpl$, at which point 
we encounter the problem with using a pseudo-Goldstone boson 
for quintessence, or ordinary inflation. 

As discussed in \cite{p97toni}
for the latter case, 
a non-renormalizable term like
$\lambda_d ^{(\Phi)}\mpl^{4-d} \Phi^d + h.c.$
will have the generic magnitude $|\lambda_d ^{(\Phi)}|\sim
M_S^4/\mpl^4$ that we discussed
before. A $Z_N$ symmetry can eliminate many such
terms, but at some order a term
$\lambda_d^{(\Phi)} \mpl^{4-N} \Phi^N + h.c.$
will eventually 
lift the potential for $\phi$. As long as $\mu\sim\mpl$, all such
terms at any order may be regarded as equally dangerous. 
Alternatively, in the spirit of
\cite{km}, we may suppose that $\lambda_d\propto
1/d!\sim e^{-d}$ for very large $d$, and ask to what order 
$\lambda_d$ must then be eliminated. The answer is $d\sim
\ln(M_S^4/V_0) \gsim 60\ln 10\sim 240$, which seems quite unreasonable.

Dual to our discussion for the modulus of $\Phi$, 
a possibility \cite{ewanpers} which has not yet been
explored (for either quintessence or ordinary inflation) is
to suppose that one has a hybrid inflation
model, where some field other than $\Phi$ is displaced from the 
minimum of the potential
and gives a constant term $V_0$ which dominates.
This would again allow $\mu\ll\mpl$, placing the non-renormalizable terms
under control, but as before the problem would be to explain
the tiny magnitude of $V_0$. 

We have yet to consider the moduli fields emerging from
string theory, which are not charged under symmetries acting on their 
phases. At present it does not seem that any of them will give 
quintessence. However, the dilaton field $s$ does look hopeful
at first sight.
At $s\gg \mpl$ its potential is supposed to be of the form
$V\propto e^{-cs}$ with $c\sim 1/\mpl$ and no corrections.
This satisfies the flatness conditions, but does not lead to viable
quintessence because the unified gauge coupling is proportional to $1/s$
and one cannot tolerate significant time-dependence for that
 coupling \cite{bin}. 
(In this article we are not considering another requirement
often imposed on quintessence models, which is that 
in the early Universe quintessence should scale with the 
radiation/matter energy density. The dilaton violates that requirement
too \cite{bin}.)

Finally, it does not help to make 
$\phi$ a condensate
rather than an elementary field.
There are actually two possibilities here. One is that 
$\phi$ exists
only {\em below} some mass scale $\Lambda\ll\mpl$, analogous to the 
situation for the Higgs in 
Technicolor extensions of the  Standard Model. This makes things much 
worse, because the effective field theory now has an ultraviolet 
cutoff $\Lambda$ and 
the natural value of the non-renormalizable coefficients $\lambda_d$ defined in
\eq{tree} is $\lambda_d\sim (\mpl/\Lambda)^{d-4}\gg 1$.
(Equivalently, the coefficients are of order 1 if we replace 
$\mpl$ by $\Lambda$). 

The opposite possibility, that 
$\phi$ exists only {\em above} some scale, 
is the one invoke \cite{bin} for the model of
\eq{dsb}. Such a behavior would be expected, for example, if $\phi$
parameterizes a flat direction in a supersymmetric theory.
However, this makes no difference at all to our discussion, because
at large
values of $\phi$ (which are required by slow roll), the theory is
weakly-coupled and $\phi$ can be
treated as a fundamental field with canonical normalization. 

{\bf 6.} In contrast with the above situation, 
ordinary inflation need not involve fine-tuning.
The basic reason is that $V$ during ordinary inflation
need not be small compared with the scale of supersymmetry breaking.
Indeed, the only theoretical  constraint is $V\leq M_S^4$,
and in fact one has $V=M_S^4$ in most models of inflation\footnote
{During ordinary 
inflation $M_S$ can differ from its present value,
but that is not a significant fact as far as fine-tuning is
concerned.}.
The value \eq{lamd} of the couplings $\lambda_d$, that can be 
achieved with supersymmetry, is then sufficient to satisfy the
flatness condition \eq{flat2} for $d>2$, provided that the model is 
constructed so that $\phi\ll \mpl$. Finally, the
mass term ($d=2$) corresponding to \eq{msquared} only marginally violates
\eq{flat2} ($m^2\sim V$ instead of $m^2\ll V$), and ways are known
that will achieve the necessary marginal reduction without 
fine-tuning.

\section*{Acknowledgements}
We would like to thank C. Csaki, A.~D.~Linde and J. Terning 
for helpful discussions.

\def\pl#1#2#3{{\it Phys. Lett. }{\bf B#1~}(19#2)~#3}
\def\zp#1#2#3{{\it Z. Phys. }{\bf C#1~}(19#2)~#3}
\def\prl#1#2#3{{\it Phys. Rev. Lett. }{\bf #1~}(19#2)~#3}
\def\rmp#1#2#3{{\it Rev. Mod. Phys. }{\bf #1~}(19#2)~#3}
\def\prep#1#2#3{{\it Phys. Rep. }{\bf #1~}(19#2)~#3}
\def\pr#1#2#3{{\it Phys. Rev. }{\bf D#1~}(19#2)~#3}
\def\np#1#2#3{{\it Nucl. Phys. }{\bf B#1~}(19#2)~#3}
\def\mpl#1#2#3{{\it Mod. Phys. Lett. }{\bf #1~}(19#2)~#3}
\def\arnps#1#2#3{{\it Annu. Rev. Nucl. Part. Sci. }{\bf #1~}(19#2)~#3}
\def\sjnp#1#2#3{{\it Sov. J. Nucl. Phys. }{\bf #1~}(19#2)~#3}
\def\jetp#1#2#3{{\it JETP Lett. }{\bf #1~}(19#2)~#3}
\def\app#1#2#3{{\it Acta Phys. Polon. }{\bf #1~}(19#2)~#3}
\def\rnc#1#2#3{{\it Riv. Nuovo Cim. }{\bf #1~}(19#2)~#3}
\def\ap#1#2#3{{\it Ann. Phys. }{\bf #1~}(19#2)~#3}
\def\ptp#1#2#3{{\it Prog. Theor. Phys. }{\bf #1~}(19#2)~#3}

\end{document}